\documentclass[%
 reprint,
 amsmath,amssymb,
 aps,
 prl,
]{revtex4-1}

\usepackage{graphicx}% Include figure files
\usepackage{dcolumn}% Align table columns on decimal point
\usepackage{bm}% bold math
\begin{document}

\preprint{APS/123-QED}

\title{A Shallow Water Analogue of the Standing Accretion Shock Instability: \\
Experimental Demonstration and Two-Dimensional Model}% Force line breaks with \\

\author{Thierry Foglizzo,$^1$ Fr\'ed\'eric Masset,$^{2,1}$ J\'er\^ome Guilet,$^{3,1}$ and Gilles Durand$^1$}
%\email{foglizzo@cea.fr}
\affiliation{%
$^1$AIM, IRFU/SAp, CEA Saclay, 91191, France\\
$^2$ICF, UNAM, 62210, Mexico\\
$^3$DAMTP Cambridge, CB3 0WA, UK
}%

\date{accepted December 13, 2011}% It is always \today, today,

\begin{abstract}
Despite the sphericity of the collapsing stellar core, the birth conditions of neutron stars can be highly non spherical due to a hydrodynamical instability of the shocked accretion flow. Here we report the first laboratory experiment of a shallow water analogue, based on the physics of hydraulic jumps. Both the experiment and its shallow water modeling demonstrate a robust linear instability and nonlinear properties of symmetry breaking, in a system which is one million times smaller and {about} hundred times slower than its astrophysical analogue.
\begin{description}
\item[PACS numbers]97.60.Bw, 95.30.Lz, 47.20.-k, 01.50.Pa
\end{description}
\end{abstract}

\maketitle

%\tableofcontents

The physics of stellar core-collapse involves complex ingredients such as nuclear physics, neutrino interactions, multidimensional hydrodynamics and general relativity. Current understanding is based on simplified formulations which make the problem numerically tractable at least, if not physically intuitive. A recent breakthrough occurred on the hydrodynamical side with the recognition of a new instability mechanism \cite{BM03, O06, F07, S08, F09, FT09, GF12} taking place in the inner 200km of the collapsing core, during a few hundred milliseconds before the explosion, while a spherical accretion shock stalls above the surface of the proto-neutron star. This shock was found to be unstable with respect to global oscillations, {with a period of a few tens of milliseconds}. An explosion is launched when sufficient neutrino energy diffusing out of the dense central region has been intercepted by the post-shock accreting matter \cite{MJ09}. {When the explosion proceeds, the large scale asymmetry resulting from this standing accretion shock instability (SASI) imposes on the neutron star a velocity kick mediated by gravitational forces \cite{S04,S06,N10, W10}, and can also affect its spin in a direction surprisingly opposite to the global rotation of the shock wave \cite{BM07, F10}. 
The pulsar spin period that could be accounted for by the spiral mode of SASI is still debated \cite{BM07,W10,F10, R11}. These effects are likely to depend on the progenitor mass and rotation rate, the physical ingredients of the numerical model, and could also be sensitive to numerical limitations such as numerical resolution. 
We shed new light on the physical nature of these processes by adopting an experimental approach for the first time.} The outcome is strikingly similar to the astrophysical results obtained through numerical simulations in an adiabatic approximation \cite{BM07}. The simplicity of the experiment in the familiar environment of shallow water physics should allow physicists to build up physical intuition about this instability.\\
\begin{figure}
\begin{center}
\includegraphics[scale=0.69]{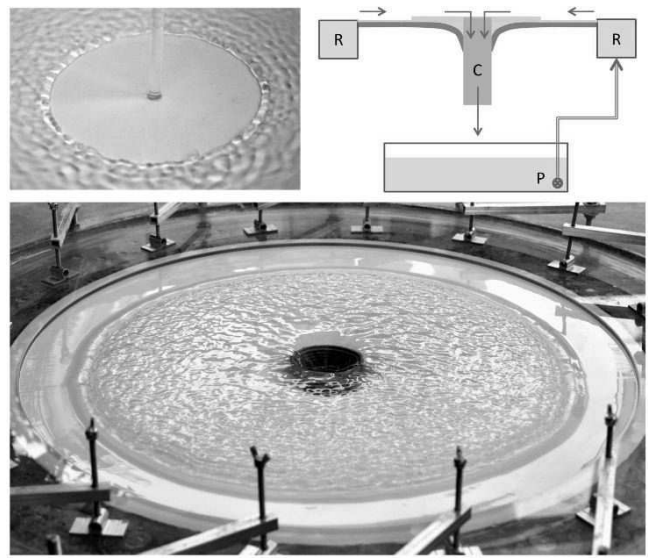}
\caption[]{Like the classical hydraulic jump in a kitchen sink (upper left), the SWASI experiment (upper right and bottom) involves a hydraulic jump associated to the deceleration of a radial flow of water. Water is injected inward from an annular injection reservoir (R) along a hyperbolic potential well, and evacuated through a vertical cylinder (C), whose walls mimic the surface of the neutron star. A pump (P) distributes collected water. The lower picture illustrates the stationary and axisymmetric character of water injection in a stable configuration.}
\label{fig1}
\end{center}
\end{figure}
{\it A Shallow Water Analogue to a Shock Instability.} The experimental set up named SWASI is designed to produce a Shallow Water Analogue of a Shock Instability. It is based on the analogy between acoustic waves in a gas and surface gravity waves in shallow water \cite{LL87}. Their velocity $c$ when the depth $H$ is smaller than their wavelength is defined by $c^2=gH$, where $g$ is the gravitational acceleration in the laboratory. The hyperbolic shape $H_{\rm grav}(r)$ of the surface on which the shallow fluid flows is a formal analogue to the gravitational potential $\Phi=gH_{\rm grav}$, dominated by the mass of the central proto-neutron star. The system of equations defining this flow in the inviscid 2D approximation can be directly compared to the equations describing an isentropic gas with an adiabatic index $\gamma=2$ \cite{N2}, where the compressible gas density plays the same role as the depth of the fluid layer in the experiment:
\begin{eqnarray}
\frac{\partial H}{\partial t}+\nabla\cdot(H v)=0,\\
\frac{\partial v}{\partial t}+(\nabla\times v)\times v+\nabla\left(\frac{v^2}{2}+{c^2}+\Phi\right)=0.\label{eq_euler}
\end{eqnarray}
The Froude number ${\rm Fr}=|v|/c$ comparing the fluid velocity $v$ to the wave velocity plays the same role as the Mach number in a gas. Shock waves are analogue to hydraulic jumps \cite{LL87}. Both can be idealized as sharp discontinuities where the mass flux and momentum flux are conserved. In contrast with the kitchen sink experiment \cite{W64} (Fig.~\ref{fig1}, upper left),  the fluid is injected inward in the SWASI experiment (Fig.~\ref{fig1}, bottom). The inner boundary is made of a vertical pipe such that water accumulating in the gravitational potential is continuously extracted by spilling over its upper edge (Fig.~\ref{fig1}, upper right). {The height of the upper edge of the inner cylinder defines a pressure threshold over which water is efficiently evacuated. It is adjusted vertically in order to choose the radius of the hydraulic jump ($r_{\rm jp}\sim20$cm). Conveniently simple, this inner boundary could be interpreted as an analogue of a cooling process such as neutrino emission close to the neutrinosphere, which decreases the energy of the gas when the density and temperature are high enough.} 
The experiment size was chosen large enough in order to minimize the effects of viscosity: the viscous drag is negligible except in the thin layer ahead of the jump.  The injection radius is 32cm, and the radius of the inner cylinder is $r_*=4$cm. The shape of the gravitational potential is described by ${\rm d}H_{\rm grav}/{\rm d}r = (5.6{\rm cm}/r)^2$.
Accreted water is pumped back into the annular reservoir through 16 pipes. A layer of sand dampens any inhomogeneity before injection.

Despite its simple set up, this experiment is expected to capture some hydrodynamical properties of the accreting gas in the equatorial plane of the stellar core \cite{BS07,F10}, particularly those observed in the adiabatic approximation \cite{BM07}. 
{The shallow water model can be scaled using the jump radius $r_{\rm jp}$ as distance unit, the free fall velocity $v_{\rm ff}\equiv (2gH_{\rm grav}^{\rm jp})^{1/2}$ as  velocity unit, and the free fall timescale $t_{\rm ff}^{\rm jp}\equiv r_{\rm jp}/v_{\rm ff}$ as time unit. The dimensionless solution depends on three parameters only \cite{N2}: the relative size of the inner boundary $r_*/r_{\rm jp}$, the pre-shock velocity $v_1/v_{\rm ff}$ and the Froude number ${\rm Fr}_1$ ahead of the shock. 
The experimental results can be scaled to astrophysical proportions by using the ratio $r_{\rm sh}/r_{\rm jp}\sim 10^6$ for distances, and the ratio ${t_{\rm ff}^{\rm sh}/t_{\rm ff}^{\rm jp}}$ for timescales:
\begin{eqnarray}
{t_{\rm ff}^{\rm sh}\over t_{\rm ff}^{\rm jp}}&\equiv& \left({r_{\rm sh}\over r_{\rm jp}}\right)^{3\over 2}\left({r_{\rm jp}gH_{\rm grav}^{\rm jp}\over GM_{\rm NS}}\right)^{1\over2}\sim 1.4\times 10^{-2},
\label{eqscaling}
\end{eqnarray}
where $M_{\rm NS}\sim 1.2M_{\rm sol}$ is the mass of the proto-neutron star and $G$ is the gravitational constant.} 

Surface gravity waves and advected vorticity perturbations present in shallow water are directly comparable to acoustic and vorticity waves in a compressible gas. These are a possible source of an unstable cycle similar to the advective-acoustic cycle in SASI \cite{O06, F07, S08, F09, FT09, GF12} and also comparable to the vortical-acoustic cycle seen in a shocked isothermal gas \cite{F02}. 

\begin{figure}
\begin{center}
\includegraphics[scale=0.42]{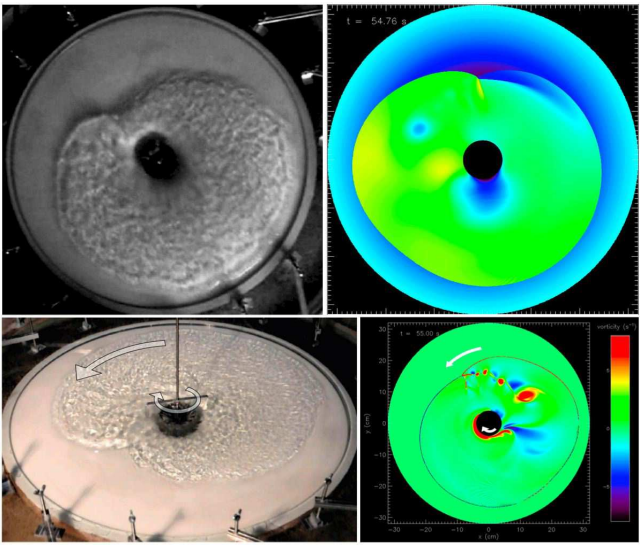}
\caption[]{The hydraulic jump governed by a spiral mode displays a rotating triple point both in the experiment (left) and in the numerical simulation of 2D shallow water equations (right). The altitude of the free surface (upper right) and the vorticity (lower right) are shown. The same shape and dynamics are observed in astrophysical simulations \cite{BM07}. The angular momentum in the accreted flow is visualized by a horizontal bar in the experiment (lower left). It spins in the direction opposite to the hydraulic jump (movie M4 \cite{N1}). The vorticity trails shown in the numerical simulation (lower right) of the experiment illustrate these counterrotating motions (movie M6 \cite{N1}).}
\label{fig2}
\end{center}
\end{figure}

\begin{figure}
\begin{center}
\includegraphics[scale=0.45]{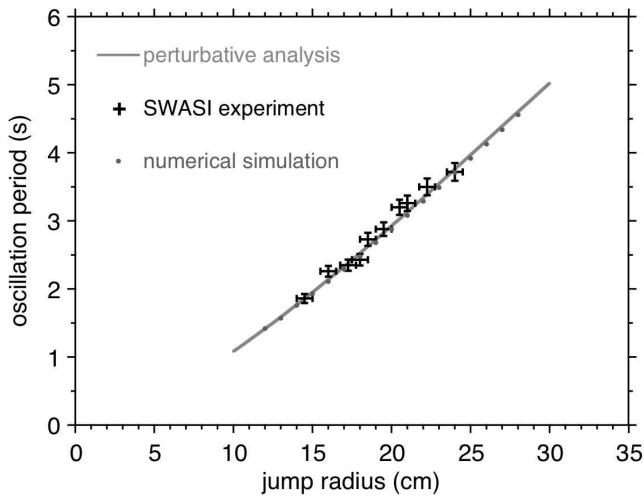}
\caption[]{The oscillation period measured in the experiment, shown here as a function of the jump radius, is well captured by both the perturbative analysis and the numerical simulation. Error bars relative to the measurement of the jump radius ($\pm5$mm) and the oscillation period ($\pm3.5\%$) are indicated. The measured flow rate is 1L/s ($\pm5\%$) and the injection slit size is 0.74mm ($\pm0.02$mm).}
\label{fig3}
\end{center}
\end{figure}

\begin{figure}
\begin{center}
\includegraphics[scale=0.27]{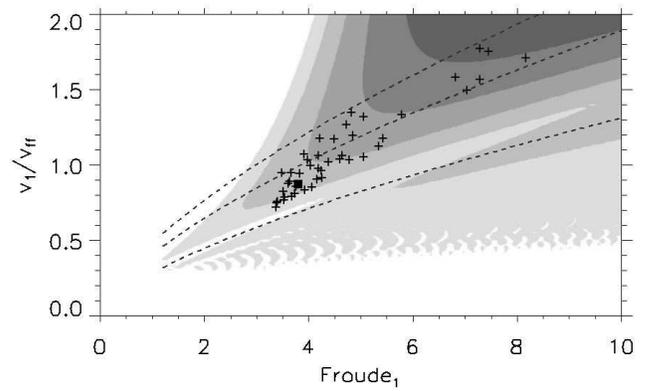}
\caption[]{ The stability of the flow depends on the strength of the hydraulic jump measured by the pre-jump
Froude number ${\rm Fr}_{1}$, and the pre-jump velocity $v_1$ measured in units of the free-fall velocity $v_{\rm ff}$ along the potential surface. 
The growth rate of the dominant $m=1$ mode, deduced from the perturbative analysis of the 2D shallow water model, {is 
displayed by contour lines for 0, 0.035, 0.07, 0.105, 0.14, 0.175 $(t_{\rm ff}^{\rm jp})^{-1}$, for a jump radius $r_{\rm jp}/r_*=4.9$ ($t_{\rm ff}^{\rm jp}=0.35{\rm s}$)}. 
The symbols '+' indicate the set of 
parameters used in experiments showing a robust $m=1$ instability, the filled square corresponds to the parameters of Fig.~\ref{fig3}. 
Dashed lines of constant flow rate (0.5, 1.5 and 2.5 L/s) {from bottom to top} correspond to a Reynolds number 408, 1224 and 2040 respectively, 
measured at the jump radius.  }
\label{fig4}
\end{center}
\end{figure}

{\it Symmetry breaking.} Indeed, as the water flux is increased, a large scale instability sets in through growing oscillations of the hydraulic jump (Fig.~\ref{fig2}, left). This instability is also observed in numerical simulations \cite{N1} of the experimental setup in the 2D shallow water approximation (Fig.~\ref{fig2}, right). The visual resemblance with astrophysical simulations is supported theoretically by the formal similarity between the set of shallow water equations and the set of adiabatic gas equations used in \cite{BM07}. A perturbative analysis of the shallow water equations \cite{N2} reveals a SASI-like instability dominated by the global mode m=1 in most of the parameter space (Fig.~\ref{fig4}). The oscillation period measured in the experiment is in excellent agreement with both the perturbative analysis and the numerical simulations (Fig.~\ref{fig3}). {Using the scaling factor in Eq.~(\ref{eqscaling}), an oscillation frequency of $3{\rm s}$ and a growth rate of $0.2{\rm s}^{-1}$ in the experiment for $v_1/v_{\rm ff}\sim1$, $r_{\rm jp}/r_*=5$ and ${\rm Fr}_1\sim  4$ would correspond to an oscillation period of $42{\rm ms}$ and a growth time of $70{\rm ms}$ on an astrophysical scale, which are comparable to the values measured in core-collapse models (e.g. Fig. 17 in \cite{S08}).}

Like 3D astrophysical simulations \cite{BM07,F10}, the experiment displays both the sloshing and the spiral modes (movies M1, M2, M3 \cite{N1}). These are easily triggered depending on the random initial perturbations, as expected since they have the same linear growth rates. However they interact during the nonlinear phase of their evolution, ultimately favoring a single right or left spiral mode even though the injected flow contains no angular momentum. This nonlinear behavior, first noted in astrophysical numerical simulations \cite{BM07}, is clearly observed in the experiment if the instability is vigorous (movies M2, M3 and M5 \cite{N1}). The shape and dynamical evolution of the hydraulic jump in the nonlinear regime is remarkably akin to the astrophysical numerical simulations \cite{BM07,BS07,F10,I08}, with the formation of a triple point when the spiral mode reaches nonlinear amplitudes (Fig. \ref{fig2}, movie M6 \cite{N1}). Another striking similarity is the fact that the angular momentum of the accreted fluid is opposite to the direction of rotation of the hydraulic jump. This is made visible in the experiment by using a light horizontal bar freely rotating on a vertical axis, with both ends bent vertically and immersed in the flowing water. The hydraulic jump and the horizontal bar systematically rotate in opposite directions (Fig.~\ref{fig2} and movies M4, M5 \cite{N1}).

{\it Effect of the bottom drag for a viscous fluid.} The Reynolds number in the experiment can be defined using the viscosity $\nu$ of water, the fluid velocity $v$ and depth $H$. It is smallest in the outer part of the experiment and inversely scales with radius:
${\cal R}e\equiv {Hv/\nu}={Q/ 2\pi\nu r}$ {where $Q$ is the injected flow rate.}
According to Fig.~\ref{fig4}, the Reynolds number at the shock radius is in the range 500-2000 within the parameters of the experiment, which supports the laminar approximation for the viscous drag \cite{M44}, proportional to $v/H^2$. This viscous drag is non negligible only in the most shallow layer ahead of the hydraulic jump. The strength of the viscous drag has been estimated experimentally in a stationary supercritical flow (Fr$>1$) by measuring the radial profile of the potential surface and the free surface of the fluid between $R=6$cm and $R=31$cm for an injection slit size of 1.3mm and a flow rate of 1.15L/s. The corresponding velocity profile was best fitted with an effective viscosity coefficient  $\bar\nu=0.03$ cm$^2/$s, {such that the viscous drag is written as an additional term ${\bar\nu}{v}/{H^2}$ in Eq.~(\ref{eq_euler}).} The perturbative analysis \cite{N2} of the 2D shallow water model assumes that this parameter is a constant.

In order to stress the analogy between the dynamics of the post-jump region in the experiment and the dynamics of the post-shock region in the astrophysical simulations, the flow parameters were chosen according to their values at the shock (velocity and Froude number) rather than at injection. Measuring the pre-jump velocity in units of the inviscid free fall velocity in Fig.~\ref{fig4}, the contour lines showing the growth rate of the instability for $\bar\nu=0.03$ cm$^2/$s are well approximated by those obtained for an inviscid flow. 

{\it Modeling approximations.} The hydraulic jump has been modeled as a simple discontinuity where energy is dissipated through a viscous roller of negligible radial extension. In reality, depending on the strength of the jump, small scale waves can be generated which may interfere with the global instability for some parameters. Intermittent behavior of hydraulic jumps are well known in the range $2.5<{\rm Fr}<4.5$, while steadier jumps are expected in the range $4.5<{\rm Fr}<9$ \cite{C73}. This latter is more typical of pre-shock Mach numbers relevant to core-collapse studies. The experiments showed possible interferences between the intermittent character of the jump and the global instability for ${\rm Fr}<3$. Fig.~\ref{fig4} shows the set of most robust experiments performed for ${\rm Fr}>3.3$.
The transition to turbulence, expected around Re$=2000$, would require a different prescription for the viscous drag (proportional to $v^2/H$). Viscosity in the horizontal plane and surface tension have been neglected.
The vertical structure of the flow has been idealized in order to obtain a 2D description of the experiment. A more accurate modeling of this vertical structure could be obtained by introducing the Coriolis and Boussinesq correction coefficients \cite{C04} for the kinetic energy and the momentum equations, in the St Venant system and in the boundary conditions. Such quantitative improvement would be at the expense of the simplicity of the analogy with gas equations. 

{\it A new tool for supernova physics.} Drawing a parallel between stellar explosions and shallow water physics is unexpected, but is deeply rooted in the universality of the laws of fluid mechanics. {The complementarity between the experimental and numerical approaches is beneficial because their inherent limitations are different. The experiment together with its idealized modeling} demonstrated the robustness of the SASI instability in a converging flow, even in a fluid without entropy gradients. This emphasizes implicitly the importance of vorticity perturbations, like in the vortical-acoustic cycle in an isothermal flow \cite{F02}.
This new tool will be used to characterize nonlinear effects determining both the saturation amplitude and the interaction between right and left spiral modes. The saturation amplitude in the inviscid approximation is expected to be governed by the parasitic growth of the Kelvin-Helmholtz instability \cite{G10}, apparent in the numerical simulations (Fig.~\ref{fig2} lower right). Furthermore, a global rotation of the experiment will be introduced in order to mimic stellar spin. This is expected to accelerate the growth of the spiral mode rotating in the same direction as the progenitor \cite{BM07,YF08} and enhance the influence of SASI on the angular momentum budget of the accretor.

This shallow water analogue of a shock instability belongs to the rare category of physical processes which are astrophysically relevant, accessible on Earth at normal temperature and pressure, and low cost. It is both a new research instrument for supernova science and an original tool for public outreach.

The design and building of the experiment benefited from the help of P. Mulet, E. Gr\'egoire, L. Rodriguez, F. Boissonnet and V. Foglizzo. This work is part of ANR funded projects Vortexplosion ANR-06-JCJC-0119 and SN2NS ANR-10-BLAN-0503. J.G. acknowledges support from STFC.

%\nocite{*}

\bibliography{foglizzo}% Produces the bibliography via BibTeX.

\end{document}